\documentclass[12pt]{article}
\pdfoutput=1       
  
\usepackage{hyperref}     
\usepackage{amsmath}       
\usepackage{amssymb} 
\usepackage{graphicx}
\usepackage{url}
\usepackage{enumerate}
\usepackage{color}
\usepackage{ulem}
 
\usepackage{float,wrapfig,color} 

\allowdisplaybreaks[1]
\def\eq#1{(\ref{#1})}
\def\s[#1\s]{\begin{align}\begin{split}#1\end{split}\end{align}}
\def\[#1\]{\begin{align}#1\end{align}}

\begin{document}

\begin{titlepage} 

\title{
\hfill\parbox{4cm}{ \normalsize YITP-24-29}\\   
\vspace{1cm} 
Usefulness of signed eigenvalue/vector\\ 
distributions of random tensors
}

\author{
Max Regalado Kloos$^1$\footnote{max.regaladokloos@gmail.com}
and 
Naoki Sasakura$^{2,3}$\footnote{sasakura@yukawa.kyoto-u.ac.jp}
\\
$^1${\small{\it Department for Physics and Astronomy, Heidelberg University, }}\\
{\small{\it Im Neuenheimer Feld 226, 69120 Heidelberg, Germany}}\\
$^2${\small{\it Yukawa Institute for Theoretical Physics, Kyoto University, }}\\
{\small{\it Kitashirakawa, Sakyo-ku, Kyoto 606-8502, Japan}}\\
$^3${\small{\it CGPQI, Yukawa Institute for Theoretical Physics, Kyoto University,}} \\
{\small{\it Kitashirakawa, Sakyo-ku, Kyoto 606-8502, Japan}}
}
  
\date{\today} 
 

\maketitle 
\begin{abstract}
Quantum field theories can be applied to compute various statistical properties of random tensors.
In particular signed distributions of tensor eigenvalues/vectors are the easiest, which
can be computed as partition functions of four-fermi theories.
Though signed distributions are different from genuine ones because of extra signs of weights,
they are expected to coincide in vicinities of ends of distributions.
In this paper, we perform a case study of the signed eigenvalue/vector distribution of 
the real symmetric order-three random tensor. 
The correct critical point and the correct end in the large $N$ limit are obtained
from the four-fermi theory, for which a method using the Schwinger-Dyson equation is very efficient.
Since locations of ends are particularly important in applications, 
such as the largest eigenvalues and the best rank-one tensor approximations, 
signed distributions are the easiest and highly useful through the Schwinger-Dyson method.
\end{abstract}
\end{titlepage}  
 
\section{Introduction}
\label{sec:introduction}
Eigenvalue distributions play important roles in random matrix models.
Wigner modeled atom Hamiltonians by random matrices and derived the celebrated semi-circle law of 
eigenvalue distributions \cite{Wigner}.
Eigenvalue distributions are used as techniques to solve matrix models \cite{Brezin:1977sv,matrix}. 
Topology changes of eigenvalue distributions give intriguing insights into the QCD 
dynamics \cite{Gross:1980he,Wadia:1980cp}.

Tensor eigenvalues/vectors\cite{Qi,lim,cart,qibook} have also many applications  in a wide range of subjects,
though the terminologies of eigenvalues/vectors are not necessarily used in the contexts.
For instance there are applications in quantum \cite{Biggs:2023mfn}/classical \cite{Evnin:2021buq} gravity, 
complexity of spin glasses \cite{pspin,pedestrians,randommat}, geometric measure of entanglement
\cite{shi,barnum}, rank-one approximation of tensors \cite{SAPM:SAPM192761164,Carroll1970}, and more \cite{qibook}. 
In these applications, it is interesting to study general properties under random tensors. 
In particular ends of eigenvalue/vector distributions of random tensors are important, since they 
provide the most extreme values typically obtained, such as the ground state energy of the spin glass 
model \cite{pspin,pedestrians,randommat}, the largest tensor eigenvalues, 
the best rank-one approximation \cite{SAPM:SAPM192761164,Carroll1970}, 
the typical value of injective norm of spiked tensor \cite{spiked}, 
and the typical amount of entanglement in terms of geometric measure \cite{estimate}.
Eigenvalue/vector distributions would also be useful in understanding the dynamics of random tensor models
\cite{Ambjorn:1990ge,Sasakura:1990fs,Godfrey:1990dt,Gurau:2009tw,Gurau:2024nzv} 

Computation of tensor eigenvalues/vectors is qualitatively different from the matrix case
and is challenging, though there are already various interesting results about their distributions for random tensors in the 
literature \cite{randommat,fyodorov1,realnum1,realnum2,Evnin:2020ddw,Gurau:2020ehg}.
Recently one of the present authors applied quantum field theoretical methods to compute 
eigenvalue/vector distributions for random tensors
\cite{Sasakura:2022zwc,Sasakura:2022iqd,Sasakura:2022axo,Sasakura:2023crd}.  
This approach using quantum field theories has some advantages: it is general, systematic, powerful, and intuitive. 
As far as random tensors are Gaussian, 
various statistical quantities of random tensors, including eigenvalue/vector distributions, can be 
reduced to computations of partition functions of zero-dimensional quantum field theories with four-interactions. 
Therefore one can apply sophisticated quantum field theoretical techniques, knowledge, and intuitions to 
compute them.

On the other hand, actual computations may become cumbersome and difficult, largely depending on 
quantum field theories obtained.  The easiest are signed distributions of eigenvalues/vectors,
which can be reduced to partition functions of quantum field theories of fermions with 
four-fermi interactions \cite{Sasakura:2022zwc}. 
In principle such partition functions  can always be exactly computed as polynomial
functions of couplings\footnote{The expression also contains a factor of an exponential of an inverse coupling.}, 
since fermion integrals \cite{zinn} are similar to taking derivatives. However, signed distributions are generally different 
from genuine distributions, because each eigenvalue/vector is weighted with a sign of a determinant of a matrix 
associated to each eigenvector.

The main purpose of this paper is to stress the usefulness of signed distributions
irrespective of the difference from genuine distributions.
We perform the case study of the signed distribution of the real eigenvalues/vectors of the real symmetric 
order-three random tensor.
We study the coincidence between the signed and genuine distributions (up to an overall sign) 
in the vicinity of the ends of the distributions, and 
derive the same critical point and the same end of the distributions in the large $N$ limit, 
for which a method using the Schwinger-Dyson equation of quantum field theory is very efficient.

The coincidence near the end between the signed and the genuine eigenvalue/vector distributions in the large $N$ limit
for the real symmetric random tensor has already been explicitly/implicitly noted 
in former references \cite{parisi,example,randommat}.
Our intention is rather to pave a path to get useful information for applications 
in various tensor problems by quantum field theories.
In fact, the usefulness of signed distributions concerning ends can generally be expected in a wide range of problems,
not being restricted to the present particular case study.
In many problems, an eigenvalue/vector equation is an equation for stationary points of a potential and the matrix mentioned
above is the Hessian of the potential on each critical point. 
In the vicinity of an end of a distribution, most of the critical points are stable or maximally unstable, meaning that
the signatures of the Hessians take a common value.  Therefore a signed and a genuine distribution should be 
almost coincident up to an overall sign in the vicinity of an end.
This implies that signed distributions, which are the easiest to compute by
quantum field theories, are indeed highly useful in various applications.

This paper is organized as follows. In Section~\ref{sec:evec} we summarize the previous results \cite{Sasakura:2022zwc}
on the signed distribution of the real eigenvalues/vectors of the real symmetric random order-three tensor. 
In Section~\ref{sec:largeN} we obtain the critical point and the end of the signed distribution 
in the large $N$ limit by two different methods, 
namely, by the Schwinger-Dyson equation of the four-fermi theory and by the analysis of Lefschetz thimbles 
(or saddle point analysis) of the signed distribution. 
We indeed obtain the same critical point and the same end as those of the genuine distribution in the large $N$ limit.
In Section~\ref{sec:smallest}  we discuss the location of an end in a different manner by
considering an equation which approximately characterizes the center of the smallest eigenvector distribution.
In Section~\ref{sec:numerics} we perform Monte Carlo simulations to crosscheck the arguments in the previous sections
by the actual data of the signed/genuine distributions.
 In Section~\ref{sec:applications}  we explain two applications of the end known in the literature, namely,
 the largest eigenvalue and the best rank-one approximation of real symmetric order-three tensors in the large $N$ limit. 
 The final section is devoted to summary and future prospects. 

\section{Eigenvalue/vector distributions of random tensors}
\label{sec:evec}
For this paper to be self-contained, in this section we summarize the results of \cite{Sasakura:2022zwc} 
about the signed distribution of 
the real eigenvalues/vectors of the real symmetric order-three random tensor.  
The eigenvector equation of such a tensor $C_{abc}\ (C_{abc}=C_{bac}=C_{bca} \in \mathbb{R}$,\ $a,b,c=1,2,\ldots,N$) 
is given by
\[
C_{abc} v_b v_c=v_a,
\label{eq:egeq}
\]
where $v_a\in \mathbb{R}$.  
Throughout this paper, repeated indices are summed over, unless otherwise stated. 
Then the distribution of the eigenvectors $v$ for a given $C$ is given by
\s[
\rho(v,C)&=\sum_{i=1}^{\# {\rm sol}(C)} \delta^N\left( v_a-v_a^{0 i}\right) \\
&=\left|\det\left[ M (v,C) \right]\right| \delta^N \left(v_a-C_{abc}v_b v_c\right),
\label{eq:rhovc}
\s] 
where  $v^{0i}\ (i=1,2,\ldots,\# {\rm sol}(C))$ are all the real solutions\footnote{We ignore the trivial solution $v=0$.} 
to \eq{eq:egeq}, 
$\delta^N(x_a)=\prod_{a=1}^N \delta(x_a)$ for a vector $x$, $|\cdot |$ is to take the absolute value, 
and $\left|\det\left[ M (v,C) \right]\right|$ is the Jacobian factor associated to the change of the arguments 
from the first line to the second. Here $M(v,C)$ is a matrix with elements,
\[
M(v,C)_{ab}=\frac{\partial}{\partial v_a} \left( v_b-C_{bcd}v_c v_d \right)=\delta_{ab}-2 C_{abc}v_c.
\]
The volume measure for the distribution \eq{eq:rhovc} as probabilities is $dv=\prod_{a=1}^N dv_a$.

When the tensor $C$ has the Gaussian (or normal) distribution, the mean distribution of $v$ is given by
taking the mean of \eq{eq:rhovc}:
\[
\rho(v)=\frac{1}{A} \int_{\mathbb{R}^{\# C}} dC\, e^{-\alpha C^2} 
\left| \det\left[M(v,C)\right] \right| \delta^N\left(v_a-C_{abc}v_bv_c\right),
\label{eq:defrho}
\]
where $\alpha$ is a positive number, $A= \int_{\mathbb{R}^{\# C}} dC\, e^{-\alpha C^2}$, 
and $\# C$ is the number of the independent elements of $C$, namely, $\# C=N(N+1)(N+2)/6$.

The real eigenvalue equation (Z-eigenvalue in the terminology of \cite{Qi}) is given by 
\[
C_{abc} w_b w_c=\zeta w_a,\  |w|=1,
\label{eq:evalue}
\]
where $\zeta \  (\in \mathbb{R}_{\geq 0})$ is an eigenvalue, $w_a \in \mathbb{R}$, and $|w|=\sqrt{w_a w_a}$. 
In fact, 
\[
\zeta=\frac{1}{|v|},
\label{eq:zeta}
\]
by comparing \eq{eq:evalue} with \eq{eq:egeq}. 
Therefore, an eigenvalue distribution can straightforwardly be computed, once $\rho(v)$ is obtained.

The expression \eq{eq:defrho} can be rewritten as a partition function of a quantum field theory of 
bosons and fermions \cite{Sasakura:2022axo}, or it can also be computed by applying a replica trick to 
a quantum field theory of fermions \cite{Sasakura:2022iqd}.
On the other hand, it is much easier to consider a quantity without taking the absolute value,
\[
\rho_{\rm signed} (v)=\frac{1}{A} \int_{\mathbb{R}^{\# C}} dC \ e^{-\alpha C^2} 
\det\left[M(v,C)\right] \delta^N\left(v_a-C_{abc}v_bv_c\right),
\label{eq:defrhosigned}
\]
because $\det\left[M(v,C)\right]$ can be represented by a pair of fermions:
$\det [M]=\int d\bar \psi d\psi \, e^{\bar \psi M \psi}$ \cite{zinn}. 
We call $\rho_{\rm signed} (v)$ a signed distribution, because the expression corresponding to
the first line of \eq{eq:rhovc} contains an extra sign,
\s[
\rho_{\rm signed} (v,C)&=\sum_{i=1}^{\# {\rm sol}(C)} {\rm Sign}\left(
\det[M(v^{0i},C)]\right) \delta^N\left( v_a-v_a^{0 i}\right),
\label{eq:rhosignvc}
\s] 
where ${\rm Sign}\left(\det[M(v^{0i},C)]\right)$ denotes the sign of $\det[M(v^{0i},C)]$.

There are no apparent connections between the two distributions \eq{eq:defrho} 
and \eq{eq:defrhosigned} (or \eq{eq:rhovc} and \eq{eq:rhosignvc}) in general. 
However, the two quantities are almost conincident up to a 
sign\footnote{Note that the overall sign of $M(v,C)$ can be flipped, because of the absolute value in \eq{eq:rhovc}.}  
near an end of a distribution, because ${\rm Sign}(\det[ M(v,C)] )$ takes the same value 
over most of the eigenvectors in the vicinity of an end. 
 The general reason for this is as follows. 
The eigenvalue equation \eq{eq:egeq} or \eq{eq:evalue} 
can be regarded as the equation for stationary points of a potential $C_{abc}v_a v_b v_c$
with a constraint $|v|=1$, and $M(v,C)$ is the Hessian on each stationary point\footnote{More precisely,
the part transverse to $v$ is the Hessian because of the constraint $|v|=1$.} .
 Since there are nearly no other stationary points below (or over) a stationary point in the vicinity of an end, 
 the signatures of the Hessian will be all positive (or negative)\footnote{In the current case, $M(v,C)$ contains 
 also the eigenvector parallel to $v$, which always gives $-1$ as the eigenvalue of $M(v,C)$. 
 This eigenvalue should be ignored in the discussion here, because of the constraint $|v|=1$. }.

As derived in \cite{Sasakura:2022zwc}, 
the quantum field theory expression of the signed distribution \eq{eq:defrhosigned} is given by
\[
\rho_{\rm signed}(v)=- 3^{(N-1)/2} \pi^{-N/2} \alpha^{N/2} |v|^{-2 N} e^{-\alpha/|v|^2}  \int d\bar \psi d\psi\, e^{S_{\rm fermion}},
\label{eq:part} 
\]
where
\[
S_{\rm fermion}=\bar \psi \cdot \psi-\frac{|v|^2}{6 \alpha} \left( \bar \psi\cdot \psi \right)^2,
\label{eq:sfermi}
\]
where the fermions are $N-1$ 
dimensional\footnote{The fermion mode parallel to $v$ is free and can
trivially be integrated out to give the overall minus sign of $\rho_{\rm signed}(v)$. There only remain the $N-1$
dimensional fermion components transverse to $v$.}, 
$\bar \psi_a,\psi_a\ (a=1,2,\ldots,N-1)$, and $\bar \psi \cdot \psi=\bar \psi_a \psi_a$.
It is not difficult to carry out the fermion integral in \eq{eq:part}, and the result is \cite{Sasakura:2022zwc}
\[
\rho_{\rm signed}(v)=-3^{1/2} \pi^{-N/2} 2^{-1+N/2} \alpha \, e^{-\alpha/|v|^2} |v|^{-N-2} U\left( 1-\frac{N}{2}, \frac{3}{2},
\frac{3 \alpha}{2 |v|^2}\right),
\]
where $U$ is the confluent hypergeometric function of the second kind. Since the expression depends only on $|v|$,
it is more convenient to express this distribution as a function of $|v|$ rather than a vector $v$. 
Multiplying the surface volume 
$2 \pi^{N/2} |v|^{N-1} /\Gamma(N/2)$ of a sphere of radius $|v|$, we obtain
\[
\rho_{\rm signed}^{\rm size}(|v|)=-\frac{2\cdot 3^{(N-1)/2} \alpha^{N/2} |v|^{-N-1}e^{-\alpha/|v|^2}}{\Gamma(N/2)}   
\int d\bar \psi d\psi\, e^{S_{\rm fermion}},
\label{eq:rhofermi}
\]
and also
\[
\rho_{\rm signed}^{\rm size}(|v|)=-\frac{3^{1/2} 2^{N/2} \alpha |v|^{-3} e^{-\alpha/|v|^2}}{\Gamma\left(N/2\right)}
U\left( 1-\frac{N}{2}, \frac{3}{2},\frac{3 \alpha}{2 |v|^2}\right).
\label{eq:rhou}
\] 
The integration measure for these distributions as probabilities is $d|v|$.

\section{The critical point and the end of the signed distribution in the large-$N$ limit }
\label{sec:largeN}
The Schwinger-Dyson method in \cite{Sasakura:2022iqd} 
has revealed that there is a critical point at $|v|_c=\sqrt{3 \alpha/(4(N-1))}$ 
in the quantum field theoretical expression of the genuine distribution. 
The critical point is a phase transition point which separates the weak coupling and the strong coupling regimes of
the quantum field theory.
In this section we will find the same critical point for the quantum field 
theoretical expression \eq{eq:part} of the signed distribution. We will also show the same by Lefschetz thimble analysis 
(or saddle point analysis) 
of the analytic expression \eq{eq:rhou}.
We will see that the critical point separates the monotonous and the infinitely oscillatory regimes of the signed distribution
in the large $N$ limit. 
We will also compute the end of the signed distribution by the two methods. It indeed agrees with the end of the 
genuine distribution in the large $N$ limit.

\subsection{Schwinger-Dyson method of the four-fermi theory}
\label{sec:sd}
The starting point of the method is to assume that the expectation values of two fermions satisfy
\s[
&\langle \bar \psi_a \psi_b \rangle=Q \delta_{ab}, \ \langle \psi_a \psi_b \rangle=\langle \bar \psi_a \bar \psi_b \rangle=0,
\label{eq:twofermion}
\s]
where the expectation values are defined by $\langle {\cal O} \rangle=
\int d\bar \psi d\psi\, {\cal O}\, e^{S_{\rm fermion}}/\int d\bar \psi d\psi \, e^{S_{\rm fermion}}$, and $Q$ will be determined later. 
\eq{eq:twofermion} is the most general form satisfying the $GL(N-1)$ symmetry, 
$\bar \psi'_a=\bar \psi_b M_{ba} , \ \psi_a' =M^{-1}_{ab} \psi_b$, of the system \eq{eq:part} (and \eq{eq:sfermi}).
From \eq{eq:twofermion} and $S_{\rm fermion}$ in \eq{eq:sfermi}, 
we obtain an effective action, 
\s[
S_{\rm eff}&=\langle S_{\rm fermion} \rangle_{\rm leading} -(N-1) \log Q \\
&= (N-1) \left(Q-g\,Q^2-\log Q\right),
\label{eq:sq}
\s]
where $g=(N-1) |v|^2/(6 \alpha)$. Here we are taking $|v|\sim 1/\sqrt{N-1}$ and 
the expectation value of $\langle S_{\rm fermion} \rangle_{\rm leading}$ 
has been computed in the leading order of $N-1$.
For instance, the expectation value of the four-fermi term has been computed by 
$\langle \bar \psi_a \psi_a \bar \psi_b \psi_b \rangle\sim \langle \bar \psi_a \psi_a  \rangle \langle \bar \psi_b \psi_b \rangle =(N-1)^2 Q^2$. More details can be found in \ref{app:sd}.

The stationary condition $\partial S_{\rm eff}/\partial Q=0$
gives a Schwinger-Dyson equation in the leading order,
\[
\frac{1}{Q} +2 g Q-1=0.
\label{eq:sd}
\]
The solution is given by
\[
Q_{\pm}=\frac{1\pm\sqrt{1-8g}}{4 g}.
\label{eq:qsol}
\]
The expression \eq{eq:qsol} shows that there is a critical point at $g_c=1/8$, or 
\[
|v|_c=\frac{1}{2} \sqrt{\frac{3 \alpha}{N-1}}\sim \frac{1}{2} \sqrt{\frac{3 \alpha}{N}},
\label{eq:criticalv}
\]
which agrees with the critical point previously found for the quantum field theory of the genuine distribution 
in \cite{Sasakura:2022iqd}.

Some discussions are needed to determine which solutions should be taken from \eq{eq:qsol}. 
At $|v|>|v|_c$, since $Q_\pm$ and therefore $S_{\rm eff}$ takes complex values, 
the partition functions for both solutions must be summed in a certain
manner to get real values for $\rho_{\rm signed} (v)$. On the other hand, at $|v|<|v|_c$, only the solution $Q_-$
should be taken, while $Q_+$ should be discarded, because one can see that $S_{\rm eff}(Q_+)$ diverges  
for $|v| \rightarrow +0$, which contradicts the actual behavior of $\rho_{\rm signed}(v)$.
In fact, in \cite{Sasakura:2022iqd}, 
a solution was chosen so as to recover the free field limit in $|v|\rightarrow +0$ (see \eq{eq:sfermi}), 
and in the present case,
the appropriate solution is $Q_-$, which takes $1$ at $|v|\rightarrow +0$. More solid argument 
will be given in the next subsection.

The critical point \eq{eq:criticalv} numerically corresponds to what is denoted by $E_\infty$ in \cite{randommat},
which they argue separates the two parameter regions where eigenvalues/vectors take finite 
indices\footnote{Denoted by $k$ in \cite{randommat}.} of $M(v,C)$ or unlimited indices in the large $N$
limit. This would be consistent with the extreme behavior of $\rho_{\rm signed}$ at $|v|>|v|_c$ in the large $N$ limit:
The exponent $S_{\rm eff}(Q_\pm)$ has an imaginary part at $|v|>|v|_c$, and,
however small it is, $\rho_{\rm signed}$ is infinitely oscillatory in the large $N$ limit.

For the purpose of applications, however, another threshold, denoted by $E_0$ in \cite{randommat},
is essential. This determines the end of the distribution, and therefore determines the best value of an optimization 
problem. This threshold exists in the region $g<g_c$, and we can determine the location by 
finding the value of $g$ satisfying $\lim_{N\rightarrow \infty} \frac{1}{N} \log  \rho_{\rm signed}(g)=0$ 
\cite{randommat}.  From the Schwinger-Dyson method above, we have 
\[
\int d\bar \psi d\psi \, e^{S_{\rm fermion}} \sim e^{(N-1) \left\{ \tilde S_{\rm eff} (Q_-)-\tilde S_{\rm eff}(Q_-(g=0),g=0)\right\}}
\label{eq:zlargen}
\]
in the large $N$ limit, where $\tilde S_\text{eff}=S_\text{eff}/(N-1)$. The exponent is subtracted by 
$\tilde S_{\rm eff}(Q_-(g=0),g=0)$, because the fermion partition function is normalized to be 1 
in the free limit ($g=0$)\footnote{This normalization is implicitly included in the formula $\det M=\int d\bar \psi d\psi\, e^{\bar \psi M \psi}$.}. By also including the coefficients in \eq{eq:rhofermi} to take the large $N$ limit,  
we obtain
\[
\rho_{\rm signed}^{\rm size}(g) \sim e^{N\left\{-\frac{1}{2}-\frac{1}{2} \log g -\frac{1}{6 g}
+\tilde S_{\rm eff} (Q_-)\right\} },
\label{eq:zerocond}
\]
where we have also inserted the actual value $\tilde S_{\rm eff}(Q_-(g=0),g=0)=1$.
The location of the end is determined by where the exponent of \eq{eq:zerocond} vanishes.
This determines
\[
g_{\rm end}\sim  0.121404\ldots,
\]
or
\[
|v|_{\rm end}=\sqrt{\frac{6 \alpha g_{\rm end}}{N-1}}\sim 0.853479 \sqrt{\frac{\alpha}{N}}.
\label{eq:vend}
\]

\subsection{Analysis of Lefschetz thimbles}
\label{sec:lefschetz}
In this section we will express \eq{eq:rhou} as a complex contour integral.
The contour can be deformed into one or a sum of steepest descent contours, called Lefschetz thimbles\footnote{
The steepest descent method in terms of Lefschetz thimbles is explained in more detail in
Section 3 of \cite{Witten:2010cx}.}, 
passing through saddle points.  What is important is that one can systematically determine the saddle points (or Lefschetz
thimbles) which should be included to express the original contour integral.  
For \eq{eq:rhou} we will find that there are two parameter regions where one saddle or two saddles should be included, 
and the transition point between the two regions agrees with \eq{eq:criticalv}. 
We also derive \eq{eq:vend} by this method.

By using the connection between the confluent hypergeometric function and the Hermite polynomial $H_\cdot(\cdot)$, 
 \eq{eq:rhou} can be rewritten as 
\[
\rho_{\rm signed}^{\rm size}(|v|) = -\frac{2^{(N+1)/2} \sqrt{\alpha}\, \Gamma((N+1)/2)}{\sqrt{\pi}\, \Gamma(N)} |v|^{-2} 
e^{-\alpha/|v|^2} H_{N-1} 
\left(\sqrt{\frac{3 \alpha}{2}}\frac{1}{|v|}\right).
\label{eq:rhohermite}
\]
Since we are interested in the regime $|v|\sim 1/\sqrt{N-1}$ from the discussions in Section~\ref{sec:sd}, let us 
introduce a new parameter $x$ by $\sqrt{N} x=\sqrt{3 \alpha /2}/|v|$. Then by using the identity 
$H_n(y)=\frac{n!}{2 \pi i} \oint_{\cal C} dz z^{-n-1} \exp(-z^2+2 y z)$,  
where ${\cal C}$ is a counterclockwise contour around the origin, the Hermite polynomial of \eq{eq:rhohermite}
can be rewritten as 
\[
H_{N-1} \left(\sqrt{\frac{3 \alpha}{2}}\frac{1}{|v|}\right)=\frac{(N-1)!\, N^{-(N-1)/2}}{2 \pi i}\oint_{\cal C} dz\,
e^{Nf(z,x)},
\label{eq:hc}
\]
where we have performed the rescaling $z\rightarrow \sqrt{N} z$ in the identity, and 
\[
f(z,x)=-z^2+2 x z - \log z.
\label{eq:f}
\]
The saddle points of \eq{eq:f} are given by
\[
z_{\pm}=\frac{1}{2}(x \pm \sqrt{-2 + x^2}),
\]
and there is a critical point at $x_c=\sqrt{2}$, which indeed agrees with $|v|_c$ of \eq{eq:criticalv}.

\begin{figure}
\begin{center}
\includegraphics[width=7cm]{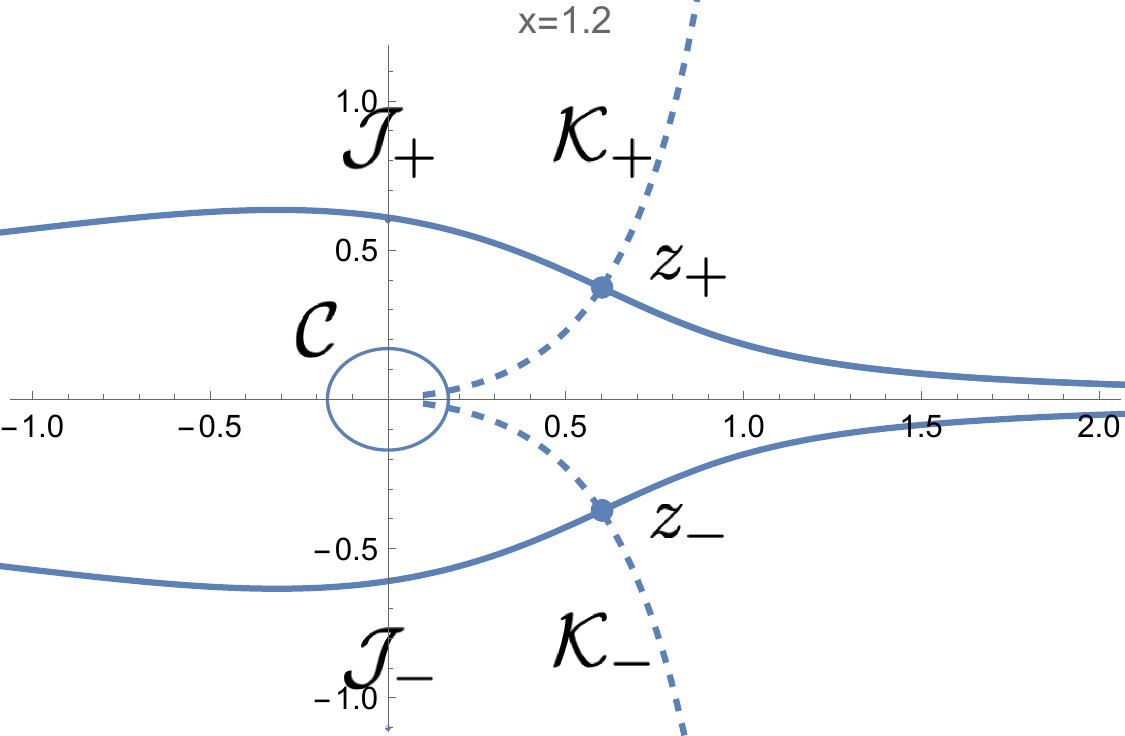}
\hfil
\includegraphics[width=7cm]{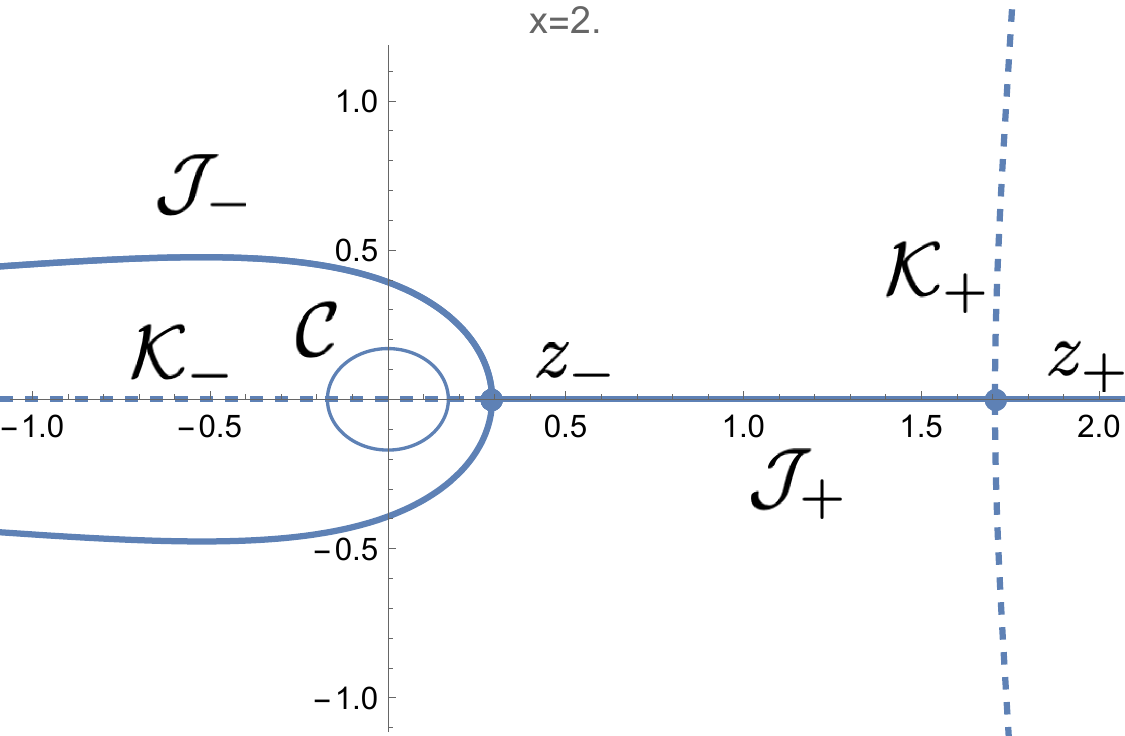}
\caption{Lefschets thimbles of \eq{eq:hc} for $x<x_c (|v|>|v|_c)$ (Left) and $x>x_c (|v|<|v|_c)$ (Right).  
Lefschetz thimbles ${\cal J}_\pm$ are shown by solid lines, while the upward flows ${\cal K}_\pm$ by dashed lines. 
The original contour ${\cal C}$ is crossed by
both ${\cal K}_\pm$ for $|v|>|v|_c$ and only by ${\cal K}_-$ for $|v|<|v|_c$.
This means that the two saddles $z_\pm$ contribute in the former case, but only $z_-$ for the latter. 
}
\label{fig:lefschetz}
\end{center}
\end{figure}

The original contour ${\cal C}$ in \eq{eq:hc} can be deformed into a sum of Lefschetz thimbles, 
which emanate from the saddle points.
A Lefschetz thimble which emanates from a saddle point $z_s$ is defined by the trajectory of the flow equation in 
a real parameter $t$, 
\[
{\cal J}_s\  :\  \frac{d z(t)}{dt}=-\overline{ \frac{\partial f(z,x)}{\partial z}},\ \ z(-\infty)=z_s, 
\]
where $\overline {X} $ denotes taking the complex conjugate of $X$. The Lefschetz thimbles which appear
in the sum can be identified by checking whether the ``upward" flow trajectory emanating from $z_s$ defined by
\[
{\cal K}_s \  :\  \frac{d z(t)}{dt}=\overline{ \frac{\partial f(z,x)}{\partial z}},\ \ z(-\infty)=z_s
\]
 crosses the original contour ${\cal C}$ in \eq{eq:hc} or not. As shown in Figure~\ref{fig:lefschetz}, 
both ${\cal K}_\pm$ cross ${\cal C}$ for $x<x_c\ (|v|>|v|_c)$, while 
only ${\cal K}_-$ does for $x>x_c\ (|v|<|v|_c)$. Therefore the original integral in \eq{eq:hc} is equal to 
the sum of those over ${\cal J}_\pm$ for $|v|>|v|_c$, while it is equal to that over ${\cal J}_-$.

For large $N$ we may well approximate $\rho_{\rm signed}^{\rm size}$ by the saddle point approximation. 
For $|v|<|v|_c$ the only saddle is real and therefore  $\rho_{\rm signed}^{\rm size}$ is monotonous.
On the other hand, $\rho_{\rm signed}^{\rm size}$ is a sum of two complex saddles at $|v|>|v|_c$. 
Looking at the form \eq{eq:hc}, it will be infinitely oscillatory in the large $N$ limit, as already discussed in Section~\ref{sec:sd}.    

One can similarly derive the same end as  \eq{eq:vend} from \eq{eq:rhohermite} and \eq{eq:hc}. In this case, the
end is located in $x>x_c$, and we obtain
\[
\rho_{\rm signed}^{\rm size}(x) \sim e^{N\left\{ f(z_-,x)-\frac{1}{2}-\frac{2 x^2}{3}\right\} }
\]
for $N\rightarrow \infty$.
The end is located at the point where the exponent vanishes. This determines
\[
x_{\rm end}=1.435\ldots,
\]
which leads to the same value as \eq{eq:vend}.
 
 \section{Asymptotic behavior of the end}
\label{sec:smallest}
In this section we will estimate the endpoint of the distribution in a different way. 
We will study the solution $v_{\rm min}$ to the equation, 
\[
\frac{1}{2}=\int_0^{v_{\rm min}} d|v| \, \left | \rho_{\rm signed}^{\rm size}(|v|)\right |.
\label{eq:onehalf}
\] 
The meaning of this equation is that $v_{\rm min}$ is roughly the center of the distribution of the smallest eigenvectors. 
In this section we will learn that the large $N$ asymptotic behavior of $v_{\rm min}$ is given by
\[
v_{\rm min}\sim \frac{1}{\sqrt{N}} \left( a_0 +\frac{a_1 \log N}{N} +\frac{a_2}{N}\right),
\label{eq:vminexp}
\]
where $a_i$ are real parameters to be tuned. This form will be used in the numerical analysis in Section~\ref{sec:numerics}.

We are going to calculate the parameters by two different methods, an approximation using the Airy function and 
the Schwinger-Dyson method. 
In both cases we want to rewrite the integral (\ref{eq:onehalf}) in the large $N$ limit in the form of
\[
B\int e^{-Nf(x)}dx
\label{integralfrom}
\]
where $B$ is depending on $N$ more mildly than exponential.  
Then by approximating $f(x) = f'(x_0)(x-x_0)$, where $f(x_0)=0$, 
we derive \eq{eq:vminexp}.

In fact the overall factor $B$ cannot fully be obtained by the leading order computation, while
$f(x)$ can. Due to this fact, the computed values of the 
parameters $a_1,a_2$ in \eq{eq:vminexp} depends on the method we take, while the value of $a_0$ and the 
asymptotic form are rather robust, as we will see in the following subsections.

\subsection{Airy Approximation}
By using the asymptotic behavior of the hermite polynomial for $n \rightarrow\infty$,
\[
e^{-\frac{X^2}{2}}\cdot H_n(X) = \pi^{\frac{1}{4}}2^{\frac{n}{2}+\frac{1}{4}}\sqrt{n!}n^{-\frac{1}{12}}\left(\text{Ai}\left(2^{\frac{1}{2}}n^{\frac{1}{6}}t\right)+O\left(n^{-\frac{2}{3}}\right)\right)
\] 
with $X=\sqrt{2n+1}+t$,
we can approximate the righthand side of \eq{eq:onehalf} with \eq{eq:rhohermite} by
\[
I = \frac{2^{N+\frac{3}{4}}\Gamma((N+1)/2)}{\pi^{\frac{1}{4}}\sqrt{3} \sqrt{\Gamma(N)}}(N-1)^{-\frac{1}{12}}e^{\frac{1-2N}{6}}\int_{t_\text{min}}^{\infty}e^{-\frac{t^2+2\sqrt{2N-1}t}{6}}\text{Ai}(\sqrt{2}(N -1)^{\frac{1}{6}}t)dt,
\label{airyapproximation}
\]
where the substitution, $\sqrt{\frac{3\alpha}{2|v|^2}}= \sqrt{2(N-1)+1}+t$, has been made. 

Following the substitution, 
we approximate the Airy function with an exponential function by using an asymptotic formula for $|\text{arg}(\sqrt{2}(N -1)^{\frac{1}{6}}t|<\pi$:
\[
\text{Ai}(z) \sim \frac{1}{2\sqrt{\pi}z^{\frac{1}{4}}}e^{-\frac{2}{3}z^{\frac{3}{2}}} \quad \text{with} \quad z = \sqrt{2}(N -1)^{\frac{1}{6}}t.
\] 
This leads us to
\[
    I \sim  \sqrt{\frac{N}{3\pi}}2^{-\frac{1}{8}}e^{\frac{1}{6}}\int_{x_\text{min}}^\infty\frac{1}{x^{\frac{1}{4}}}e^{-\frac{N}{3}[\frac{1}{2}x^2+\sqrt{2}x+2^{\frac{7}{4}}x^{\frac{3}{2}}-\frac{3}{2}\log(2)+1]}dx
    \label{expintairy}
\]
for large $N$,
where we have further performed the substitution $t=\sqrt{N}x$, and have used $\Gamma(m)=(m-1)!$ for positive 
integers $m$ and the Stirling's formula to approximate the factorials.

By approximating the function in the exponent near its root point $x_0=0.0208$, we get a very simple integral:
\[
    I \sim \sqrt{\frac{N}{3\pi}}2^{-\frac{1}{8}}e^{\frac{1}{6}}\int_{y_\text{min}}^\infty\frac{1}{(y+x_0)^\frac{1}{4}}e^{-0.721Ny}dy, 
    \label{intairyfinalapprox}
\]
where we have introduced $y = x-x_0$.
Due to the fact that the integrand is only relevant for small $y$ for large $N$, 
we can assume $(y+x_0)^\frac{1}{4}\sim x_0^\frac{1}{4}$.

By explicitly performing the integration \eq{intairyfinalapprox} and setting (\ref{eq:onehalf}),
we obtain
\[
v_\text{min} \sim \sqrt{\frac{\alpha}{N}}\left(0.853 + 0.413 \frac{\log N}{N}- 0.5710\frac{1}{N} \right)
\label{vminairy},
\]
asymptotically for large $N$. The first coefficient indeed agrees with $|v|_\text{end}$ in \eq{eq:vend}. 

\subsection{Schwinger-Dyson method}
In this subsection, we study the behavior \eq{eq:vminexp} for the expression derived from the 
Schwinger-Dyson method. By putting \eq{eq:zlargen} into \eq{eq:rhofermi}, we obtain
\[
\rho_{\rm signed}^{\rm size}(|v|)=-\frac{2\cdot 3^{(N-1)/2} \alpha^{N/2} |v|^{-N-1}}{\Gamma(N/2)} e^{-\alpha/|v|^2}  
e^{(N-1) \left\{ \tilde S_{\rm eff} (Q_-)-\tilde S_{\rm eff}(Q_-(g=0),g=0)\right\}}
\label{eq:rsignedsd}
\]
for $|v|<|v|_c$. 
On the other hand, the large $N$ limit of the genuine distribution was obtained in \cite{Sasakura:2022iqd}
by the Schwinger-Dyson method. With $v^2 = x\frac{3\alpha}{N-1}$, it is given by 
\[
\rho_\text{genuine}^\text{size}(|v|) = \frac{2\cdot 3^{(N-1)/2}\alpha^{N/2}|v|^{-(N+1)}}{\Gamma(N/2)}e^{-\alpha/|v|^2}e^{\frac{(N-1)}{4}\left(2+\log(16)+\frac{1-\sqrt{1-4x}}{x}-4\log(1-\sqrt{1-4x})+4\log x\right)-(N-1)},
\label{rhodyson}
\]
for $x<1/4$. In fact one can check that \eq{eq:rsignedsd} and \eq{rhodyson} are identical up to the minus sign.

The integral of (\ref{eq:onehalf}) with \eq{rhodyson} can be transformed into
\[
I = \frac{1}{2}\sqrt{\frac{N}{3\pi}}\int_0^{x_\text{min}}e^{-Nf(x) + g(x)}dx
\label{formintdyson}
\]
by substitution, where $f(x)$ and $g(x)$ equal
\[
\label{f(x)dyson}
f(x) &= -\frac{1}{2}\log x + \frac{1}{4x}\left(\frac{1}{3} + \sqrt{1-4x}\right) + \log (1-\sqrt{1-4x}) - \frac{3}{2}\log 2 ,\\
\label{g(x)dyson}
g(x) &= -2\log x + \frac{1}{4x}\left(\frac{1}{3} + \sqrt{1-4x}\right) + \log (1-\sqrt{1-4x}) - \log 2
\]
By approximating $f(x)$ near its root point $x_0=0.2428$, we can simplify the integral for large $N$:
\[
I \sim \frac{1}{2}\sqrt{\frac{N}{3\pi}}e^{g(x_0)}\int_0^{x_\text{min}}e^{-N[f'(x_0) - \frac{1}{N}g'(x_0)](x-x_0)}dx
 \sim \frac{1}{2}\sqrt{\frac{N}{3\pi}}e^{g(x_0)}\int_0^{x_\text{min}}e^{-Nf'(x_0)(x-x_0)}dx.
\label{intdyson}
\]
The solution of the equation (\ref{eq:onehalf}) for this integral is:
\[
v_\text{min} \sim \sqrt{\frac{\alpha}{N}}\left(0.853 + 0.412 \frac{\log N}{N}- 0.487\frac{1}{N} \right).
\label{vmindyson}
\]

\section{Numerical support}
\label{sec:numerics}
In this section we will show some numerical evidences which support 
the discussions in the previous sections.
The procedure of the Monte Carlo (MC) simulations is the same as in the previous works \cite{Sasakura:2022zwc,
Sasakura:2022iqd,Sasakura:2022axo,Sasakura:2023crd}. 
In more detail, we repeat the following sampling processes, and 
the number of repetitions is denoted by $N_{\rm samp}$:
\begin{itemize}
\item 
Randomly generate $C$ by $C_{abc}={\cal N}(0,1)/\sqrt{d(a,b,c)}$. Here ${\cal N}(0,1)$ denotes 
the normal distribution of mean value zero and standard deviation 1, which corresponds
to\footnote{Since $\alpha$ determines only the overall scale, this particular choice does not cause the
loss of generality.} $\alpha=1/2$. 
$d(a,b,c)$ is the degeneracy\footnote{This factor is because we use the weight $e^{-\alpha C_{abc} C_{abc}}$ for
the symmetric tensor $C$, where 
$C_{abc}C_{abc}=\sum_{a\leq b\leq c=1}^N d(a,b,c)\, C_{abc}C_{abc}$.}: $d(a,a,a)=1,\ d(a,b,b)=d(b,a,b)=d(b,b,a)=3,\ d(a,b,c)=6, \hbox{ for } a \neq b\neq c\neq a$. 

\item
Solve the eigenvector equation \eq{eq:egeq} for a generated $C$. Store all the real eigenvectors as sets 
$\left(|v|, {\rm\ Signature\ of\ } M(v,C)\right)$.
\end{itemize}

We took $N_{\rm samp}=10000$ for $6\leq N \leq 11$, $N_{\rm samp}=3000$ for $N=12$,
and $N_{\rm samp}=1000$ for $N=13,14$. 
The machine was a workstation with a Xeon W2295 (3.0GHz, 18 cores), 128GB DDR4 memory, and Ubuntu 20 as OS.
The eigenvector equation \eq{eq:egeq} was solved by the NSolve command of Mathematica.

\begin{figure}
\begin{center}
\includegraphics[width=5cm]{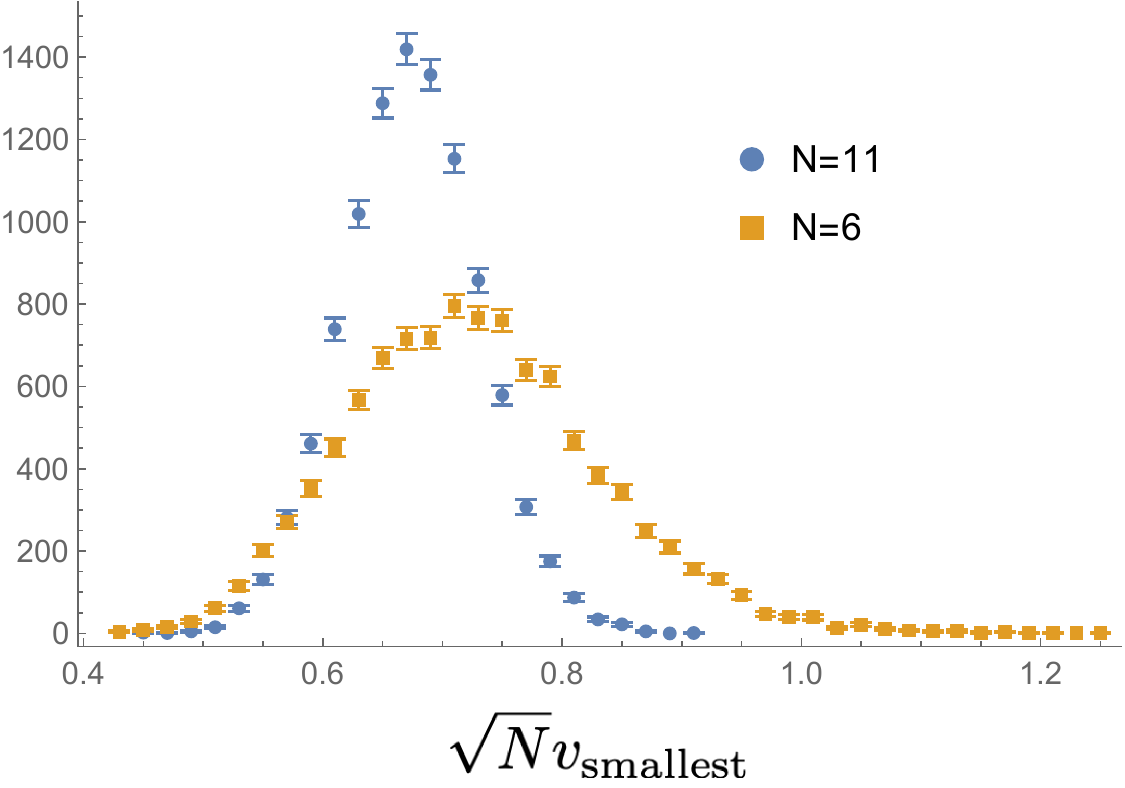}
\hfil
\includegraphics[width=5cm]{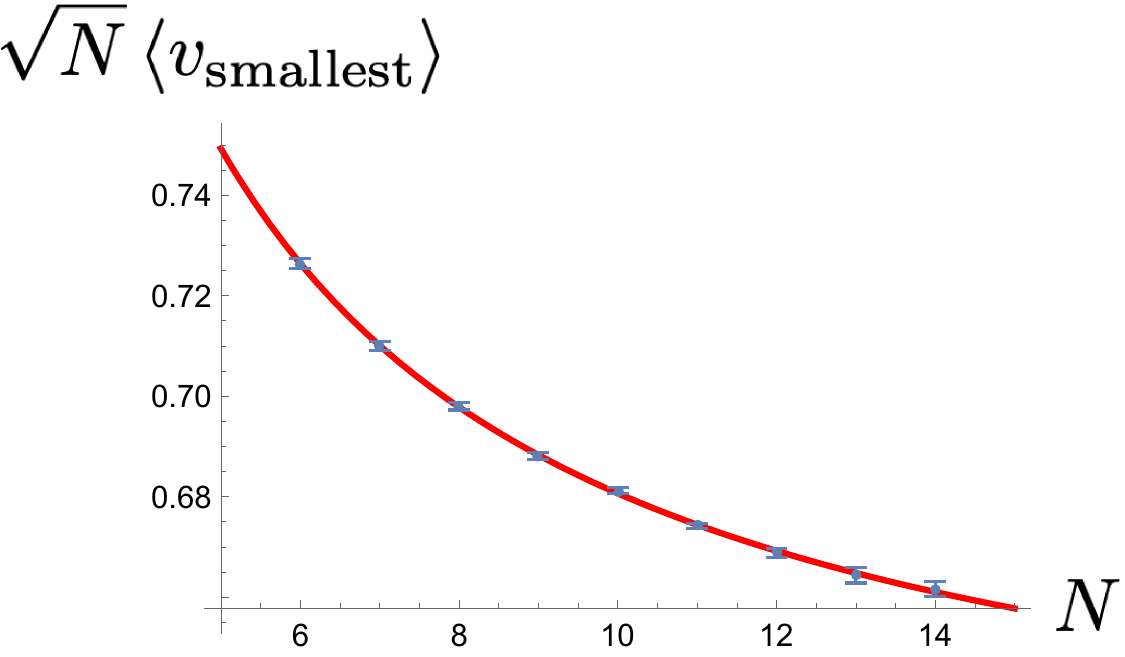}
\hfil
\includegraphics[width=5cm]{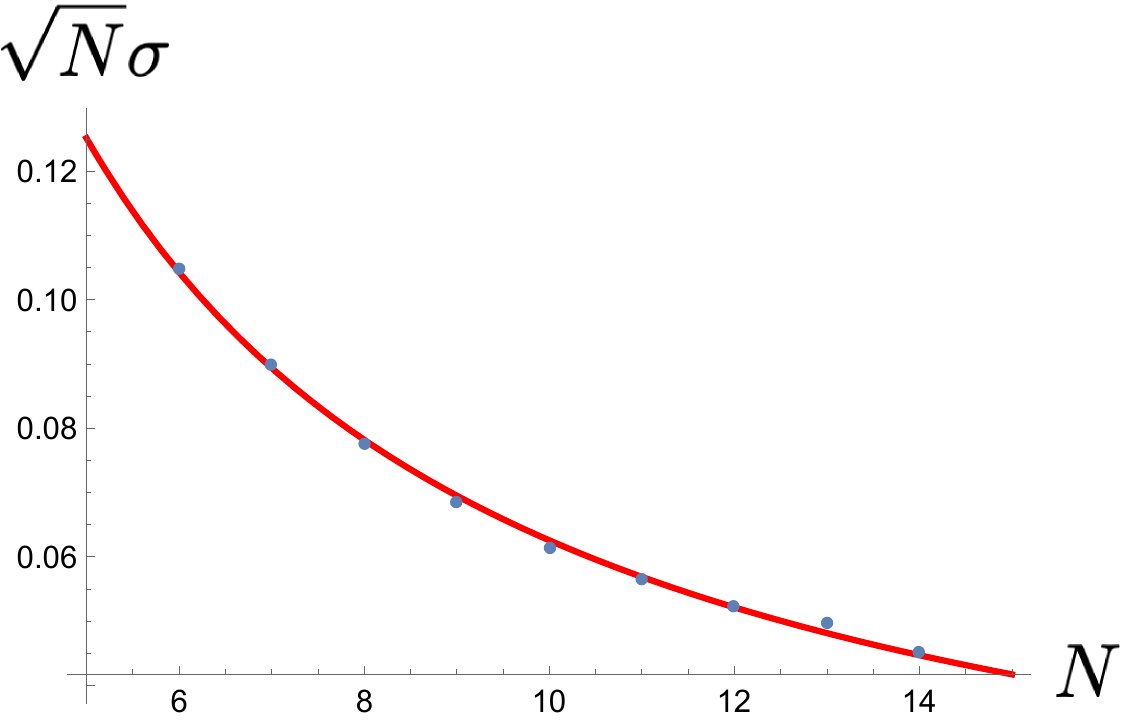}
\caption{The size distribution of the smallest eigenvectors for $N=6,11$ (Left). The horizontal axis is rescaled as $\sqrt{N} 
v_{\rm smallest}$. Mean values of the size distributions $\sqrt{N} \langle v_{\rm smallest} \rangle$ 
are plotted against $N$ (Middle). The data can be fitted well with
$\sqrt{N} \langle v_{\rm smallest} \rangle\sim 0.612\pm 0.011  + (0.00 \pm 0.10) \log N/N+ (0.69 \pm 0.11)/N$. 
The standard deviation $\sigma$ of the distribution can well be fitted with $\sqrt{N} \sigma \sim 0.63/N$ (Right).}
\label{fig:smallest}
\end{center}
\end{figure}

Suppose a tensor $C$ is given. Then it has the smallest eigenvector, and let us denote the size by $v_{\rm smallest}(C)$. 
Collecting $v_{\rm smallest}(C)$ for randomly generated  $C$, we 
obtain its distribution. Figure~\ref{fig:smallest} shows some results of MC on $v_{\rm smallest}$. 
In the left figure the distributions of the sizes of the smallest eigenvectors are shown for $N=6,11$.
In the middle figure the dependence on $N$ of the mean values $\langle v_{\rm smallest}\rangle$
of the distributions are shown, and it is fitted with $N^{-1/2} \left(a_0+a_1 N^{-1} \log(N) +a_2 N^{-1}\right)$,
motivated by the discussions in Section~\ref{sec:smallest}.
The fitted value $a_0=0.612\pm 0.011$ indeed agrees well with  
$\sqrt{N} |v|_{\rm end}\sim 0.603501\ (\alpha=1/2)$ expected  from \eq{eq:vend}.
The distribution becomes sharper as $N$ becomes larger: the standard deviation $\sigma$ of the distribution 
behaves as $\sqrt{N} \sigma \sim N^{-1}$. 
Thus the MC result implies that there is a sharp end in the large $N$ limit.

\begin{figure}
\begin{center}
\includegraphics[width=7cm]{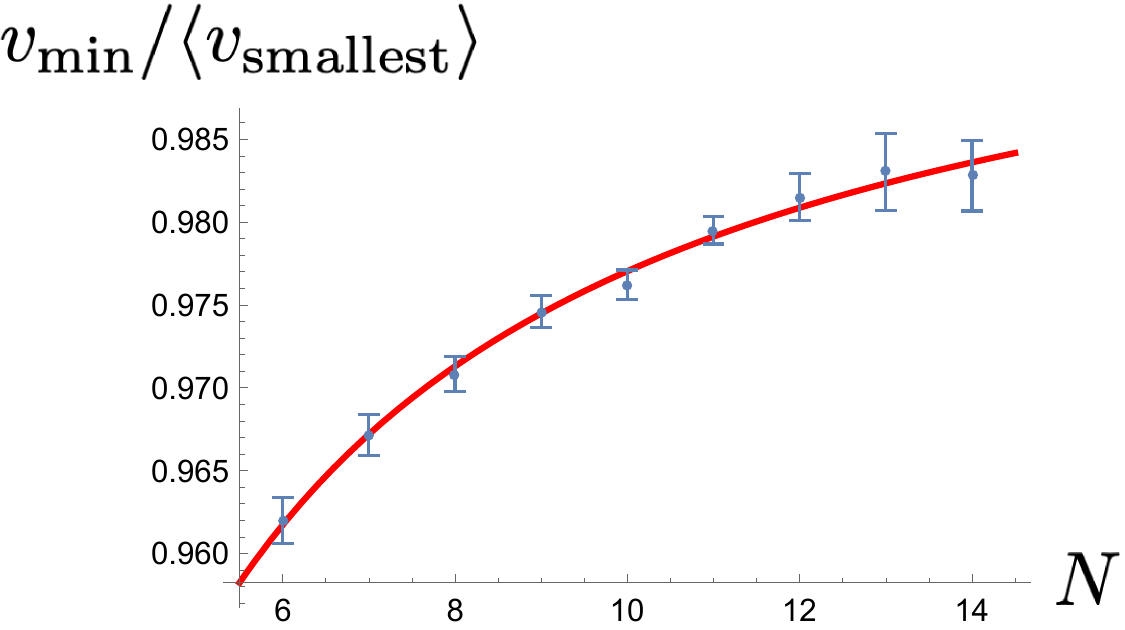}
\caption{
Left: The ratio $v_{\rm min}/\langle v_{\rm smallest}\rangle$ against $N$. It can be fitted with 
$1.000\pm 0.002-(0.23\pm 0.02)/N$.}. 
\label{fig:vmin}
\end{center}
\end{figure}

Next we want to check how well $v_{\rm min}$ represents $\langle v_{\rm smallest}\rangle$.
The values of $v_{\rm min}$ can be computed by numerically solving \eq{eq:onehalf} with \eq{eq:rhou}.
Figure~\ref{fig:vmin} shows the $N$ dependence of the ratios between $v_{\rm min}$ and 
$\langle v_{\rm smallest}\rangle$ of the MC. 
The ratio is indeed different from 1, but the difference is small and becomes smaller as $1/N$. 

\section{Applications}
\label{sec:applications}
There are various optimization/physics problems which tensor eigenvalues/vectors can be applied to,
as explained in Section~\ref{sec:introduction}. In such a problem,
an eigenvalue/vector equation is the stationary condition of a potential to be optimized/minimized, and 
it is often the case that the smallest eigenvector is the best configuration for the optimization/minimization. 
Then an end of an eigenvalue/vector distribution for random tensors gives the typical best 
optimized/minimal value/configuration. 
Moreover, the matrix $M(v,C)$ is the Hessian of the potential, meaning that 
the location of an end can be derived from the signed distribution.
In this section we explain two simple applications in the literature \cite{qibook,SAPM:SAPM192761164,Carroll1970}.

An application of the end of the distribution \eq{eq:vend} is to the largest real eigenvalue of the real symmetric 
order-three random tensor in the large $N$ limit. 
From \eq{eq:zeta}, the largest eigenvalue is determined by the smallest eigenvector. 
Therefore in the large $N$ limit the largest eigenvalues converge to  
\[
\zeta_{\rm largest}=\frac{1}{|v|_{\rm end}}\sim1.17167 \sqrt{\frac{N}{\alpha}}.
\label{eq:largesteg}
\]

Another application is to the best rank-one approximation. The best real symmetric  rank-one approximation of a 
real symmetric order-three tensor $C$ is defined by finding a real vector $\phi$, whose tensor product 
maximally approximates
$C$, namely,
\[
\min_{\phi} \left(C_{abc}-\phi_a \phi_b \phi_c\right)\left(C_{abc}-\phi_a \phi_b \phi_c\right).
\]
Considering the stationary condition, we obtain
\[
C_{abc}\phi_b \phi_c -|\phi|^4 \phi_a=0,
\label{eq:eqphi}
\]
or equivalently,
\[
C_{abc}w_b w_c -\zeta w_a=0
\]
with $\zeta=|\phi|^3$ and $w_a = \phi_a/|\phi|$. As defined in \eq{eq:evalue}, $\zeta$ is an eigenvalue of $C$.
For a solution $\phi$ satisfying \eq{eq:eqphi} we obtain
\[
\left(C_{abc}-\phi_a \phi_b \phi_c\right)\left(C_{abc}-\phi_a \phi_b \phi_c\right)=C^2-\zeta^2.
\]
Therefore the best rank-one approximation is provided by the $w$ 
which has the largest eigenvalue, or the smallest eigenvector $v$. 
By using \eq{eq:largesteg} and $ \langle C^2 \rangle_C=-\frac{\partial}{\partial \alpha}  \log \int dC e^{-\alpha C^2}=
N(N+1)(N+2)/(12 \alpha)$, the ratio between the size of $C$ and that of the
best approximating rank-one tensor $\phi_a^{\rm best}\phi_b^{\rm best}\phi_c^{\rm best}$ is given by
\[
\frac{\sqrt{\phi_a^{\rm best}\phi_b^{\rm best}\phi_c^{\rm best}\phi_a^{\rm best}\phi_b^{\rm best}\phi_c^{\rm best}} }{\sqrt{\langle C^2 \rangle_C}}
=\frac{\zeta_{\rm largest} }{\sqrt{\langle C^2 \rangle_C}}\sim \frac{4.0588}{N}.
\]  

\section{Summary and future prospects}
In this paper, we have argued the usefulness of signed distributions by performing in detail a case study
of the signed distribution of the real eigenvalues/vectors of the real symmetric order-three random tensor.
The usefulness essentially comes from the fact that the signed and the genuine distributions are almost coincident
near the ends of the distributions. In particular they have the same critical point and the end in the large $N$ limit. 
We have seen that the Schwinger-Dyson method is the most efficient way to get these values.
We have explained a few applications of the location of the end.

Though the argument of this paper is based on a detailed case study of a particular signed 
eigenvalue/vector distribution, similar usefulness can be expected for a wide range of signed eigenvalue/vector distributions
of various tensor problems. 
The main requirement for the usefulness is that eigenvectors are stationary points of a potential 
which should be optimized/minimized. In such a case, the matrix $M(v,C)$ is Hessian on a stationary point 
and the Hessians of most of the stationary points in the vicinity of an end will take all positive (or negative) 
signatures, since they are stable (or maximally unstable) points. 
Then $\det M(v,C)$ is constant among these points and the signed and the genuine distributions almost coincide 
up to an overall sign. 
Since locations of ends are essential for some applications, the signed distribution is not only the easiest to 
compute through a four-fermi theory but also very useful for applications.
We expect that signed distributions of various optimization problems could be rewritten as 
partition functions of four-fermi theories, from which we can efficiently extract useful information  
by the Schwinger-Dyson method.

\vspace{.3cm}
\section*{Acknowledgements}
N.S. is supported in part by JSPS KAKENHI Grant No.19K03825.  We would like to thank 
N.~Delporte, O.~Evnin, and L.~Lionni for stimulating discussions. 

\appendix
\def\thesection{Appendix  \Alph{section}}

\section{The Schwinger-Dyson equation}
\label{app:sd}
A full account of the Schwinger-Dyson method used here and in the previous paper \cite{Sasakura:2022iqd} 
can be found in an appendix of \cite{Sasakura:2022iqd}. 
In this appendix, on the other hand, we specifically consider \eq{eq:sfermi} and directly derive \eq{eq:sd}.

The starting point is the vanishing of the integral of the total derivative, 
\[
\int d\bar \psi d\psi\ \frac{\partial}{\partial \bar \psi_b} \left( \bar \psi_a e^{S_{\rm fermion}}\right)=0,
\]
which trivially holds for a fermionic integration \cite{zinn}.
By explicitly performing the derivative in the integrand, we obtain
\[
\int d\bar \psi d\psi \left( \delta_{ab}- \bar \psi_a \psi_b + \frac{|v|^2}{3 \alpha}  \bar \psi_a \psi_b \bar\psi \cdot \psi \right)e^{S_{\rm fermion}}
=0,
\] 
which leads to 
\[
\delta_{ab}-\langle  \bar \psi_a \psi_b \rangle +  \frac{|v|^2}{3 \alpha}  \langle \bar \psi_a \psi_b \bar\psi_c \psi_c  \rangle=0.
\label{eq:sdcor}
\]
In the leading order of $N-1$, the last term can be approximated by
\s[
\langle \bar \psi_a \psi_b \bar \psi_c \psi_c \rangle \sim \langle \bar \psi_a \psi_b \rangle \langle \bar \psi_c \psi_c\rangle
= (N-1) Q^2 \delta_{ab},
\s]
where we have used \eq{eq:twofermion} and have ignored connected four fermion expectation values. 
Putting this and \eq{eq:twofermion} into \eq{eq:sdcor}
and taking the highest order of $N-1$ , we obtain \eq{eq:sd}.


\vspace{.3cm} 

\begin{thebibliography}{}

\bibitem{Wigner} E.~P.~Wigner, ``On the Distribution of the Roots of Certain Symmetric Matrices," Annals of Mathematics {\bf 67} (2), 325-327 (1958) https://doi.org/10.2307/1970008.

\bibitem{Brezin:1977sv}
E.~Brezin, C.~Itzykson, G.~Parisi and J.~B.~Zuber,
``Planar Diagrams,''
Commun. Math. Phys. \textbf{59}, 35 (1978)
doi:10.1007/BF01614153

\bibitem{matrix}
B.~Eynard, {\it Counting surfaces : CRM Aisenstadt chair lectures},  Prog. Math. Phys. {\bf 70}, 
Birkh\"auser, 2016.

\bibitem{Gross:1980he}
D.~J.~Gross and E.~Witten,
``Possible Third Order Phase Transition in the Large N Lattice Gauge Theory,''
Phys. Rev. D \textbf{21}, 446-453 (1980)
doi:10.1103/PhysRevD.21.446

\bibitem{Wadia:1980cp}
S.~R.~Wadia,
``$N$ = Infinity Phase Transition in a Class of Exactly Soluble Model Lattice Gauge Theories,''
Phys. Lett. B \textbf{93}, 403-410 (1980)
doi:10.1016/0370-2693(80)90353-6

\bibitem{Qi}
L.~Qi, ``Eigenvalues of a real supersymmetric tensor," Journal of Symbolic Computation \textbf{40},
1302-1324 (2005).

\bibitem{lim}
L.H.~Lim, ``Singular Values and Eigenvalues of Tensors: A Variational Approach," in Proceedings of the IEEE International Workshop on Computational Advances in Multi-Sensor Adaptive Processing (CAMSAP '05), Vol. 1 (2005), pp. 129--132.

\bibitem{cart}
D.~Cartwright and B.~Sturmfels, ``The number of eigenvalues of a tensor," Linear algebra and its
applications \textbf{438}, 942-952 (2013).

\bibitem{qibook}
L.~Qi, H.~Chen, Y,~Chen,  {\it Tensor Eigenvalues and Their Applications}, Springer, Singapore, 2018.

\bibitem{Biggs:2023mfn}
A.~Biggs, J.~Maldacena and V.~Narovlansky,
``A supersymmetric SYK model with a curious low energy behavior,''
[arXiv:2309.08818 [hep-th]].

\bibitem{Evnin:2021buq}
O.~Evnin,
``Resonant Hamiltonian systems and weakly nonlinear dynamics in AdS spacetimes,''
Class. Quant. Grav. \textbf{38}, no.20, 203001 (2021)
doi:10.1088/1361-6382/ac1b46
[arXiv:2104.09797 [gr-qc]].

\bibitem{pspin}
A. Crisanti and H.-J. Sommers, ``The spherical p-spin interaction spin glass model: the statics", Z. ~Phys. 
\textbf{B 87}, 341 (1992).

\bibitem{pedestrians}
T.~Castellani and A.~Cavagna, ``Spin-glass theory for pedestrians'', 
J.~Stat.~Mech.: Theo.~Exp. {\bf 2005}, P05012
[arXiv: cond-mat/0505032].  

\bibitem{randommat}
A.~Auffinger, G.B.~Arous, and J.~\v{C}ern\'{y}, ``Random Matrices and Complexity of Spin Glasses", 
Comm. Pure Appl. Math., \textbf{66}, 165-201 (2013)
doi.org/10.1002/cpa.21422
[arXiv:1003.1129 [math.PR]].


\bibitem{shi}
A.~Shimony, ``Degree of entanglement," 
Annals of the New York Academy of Sciences, \textbf{755} (1), 675–679, (1995)
doi:10.1111/j.1749-6632.1995.tb39008.x

\bibitem{barnum}
H.~Barnum and N.~Linden, ``Monotones and invariants for multi-particle quantum states," 
Journal of Physics A: Mathematical and General, \textbf{34} (35), 6787, (2001)
doi:10.1088/0305-4470/34/35/305

\bibitem{SAPM:SAPM192761164}
F.~L. Hitchcock, ``The expression of a tensor or a polyadic as a sum of
  products,'' \href{http://dx.doi.org/10.1002/sapm192761164}{{\em Journal of
  Mathematics and Physics} {\bfseries 6} no.~1-4, (1927) 164--189}.
  \url{http://dx.doi.org/10.1002/sapm192761164}.

\bibitem{Carroll1970}
J.~D. Carroll and J.-J. Chang, ``Analysis of individual differences in
  multidimensional scaling via an n-way generalization of ``eckart-young''
  decomposition,'' \href{http://dx.doi.org/10.1007/BF02310791}{{\em
  Psychometrika} {\bfseries 35} no.~3, (Sep, 1970) 283--319}.
  \url{https://doi.org/10.1007/BF02310791}.

\bibitem{spiked}
A.~Perry,  A.~S.~Wein, A.~S.~Bandeira, ``Statistical limits of spiked tensor models,"
Ann. Inst. H. Poincaré Probab. Statist. \textbf{56} (1), 230-264, (2020). 
doi:10.1214/19-AIHP960
[arXiv:1612.07728 [math.PR]].

\bibitem{estimate}
K.~Fitter, C.~Lancien, I.~Nechita,
``Estimating the entanglement of random multipartite quantum states,"
[arXiv:2209.11754 [quant-ph]].


\bibitem{Ambjorn:1990ge}
J.~Ambjorn, B.~Durhuus and T.~Jonsson,
``Three-dimensional simplicial quantum gravity and generalized matrix models,''
Mod. Phys. Lett. A \textbf{6}, 1133-1146 (1991)
doi:10.1142/S0217732391001184

\bibitem{Sasakura:1990fs}
N.~Sasakura,
``Tensor model for gravity and orientability of manifold,''
Mod. Phys. Lett. A \textbf{6}, 2613-2624 (1991)
doi:10.1142/S0217732391003055

\bibitem{Godfrey:1990dt}
N.~Godfrey and M.~Gross,
``Simplicial quantum gravity in more than two-dimensions,''
Phys. Rev. D \textbf{43}, R1749(R) (1991)
doi:10.1103/PhysRevD.43.R1749

\bibitem{Gurau:2009tw}
R.~Gurau,
``Colored Group Field Theory,''
Commun. Math. Phys. \textbf{304}, 69-93 (2011)
doi:10.1007/s00220-011-1226-9
[arXiv:0907.2582 [hep-th]].

\bibitem{Gurau:2024nzv}
R.~Gurau and V.~Rivasseau,
``Quantum Gravity and Random Tensors,''
[arXiv:2401.13510 [hep-th]].

\bibitem{fyodorov1}
Y.V.~Fyodorov, 
``Topology trivialization transition in random non-gradient autonomous ODEs on a sphere,"
J. Stat. Mech. (2016) 124003
doi:10.1088/1742-5468/aa511a
[arXiv:1610.04831 [math-ph]].

\bibitem{realnum1}
P.~Breiding, ``The expected number of eigenvalues of a real gaussian tensor," SIAM Journal on Applied 
Algebra and Geometry \textbf{1}, 254-271 (2017).

\bibitem{realnum2}
P.~Breiding, ``How many eigenvalues of a random symmetric tensor are real?," Transactions of the
American Mathematical Society \textbf{372}, 7857-7887 (2019).

\bibitem{Evnin:2020ddw}
O.~Evnin,
``Melonic dominance and the largest eigenvalue of a large random tensor,''
Lett. Math. Phys. \textbf{111}, 66 (2021)
doi:10.1007/s11005-021-01407-z
[arXiv:2003.11220 [math-ph]].

\bibitem{Gurau:2020ehg}
R.~Gurau,
``On the generalization of the Wigner semicircle law to real symmetric tensors,''
[arXiv:2004.02660 [math-ph]].

\bibitem{Sasakura:2022zwc}
N.~Sasakura,
``Signed distributions of real tensor eigenvectors of Gaussian tensor model via a four-fermi theory,''
Phys. Lett. B \textbf{836}, 137618 (2023)
doi:10.1016/j.physletb.2022.137618
[arXiv:2208.08837 [hep-th]].

\bibitem{Sasakura:2022iqd}
N.~Sasakura,
``Real tensor eigenvalue/vector distributions of the Gaussian tensor model via a four-fermi theory,''
PTEP \textbf{2023}, no.1, 013A02 (2023)
doi:10.1093/ptep/ptac169
[arXiv:2209.07032 [hep-th]].

\bibitem{Sasakura:2022axo}
N.~Sasakura,
``Exact analytic expressions of real tensor eigenvalue distributions of Gaussian tensor model for small $N$,''
J. Math. Phys. \textbf{64}, 063501 (2023)
doi:10.1063/5.0133874
[arXiv:2210.15129 [hep-th]].

\bibitem{Sasakura:2023crd}
N.~Sasakura,
``Real eigenvector distributions of random tensors with backgrounds and random deviations,''
PTEP \textbf{2023}, no.12, 123A01 (2023)
doi:10.1093/ptep/ptad138
[arXiv:2310.14589 [hep-th]].

\bibitem{zinn}
J. Zinn-Justin, {\it Quantum Field Theory and Critical Phenomena}, (Clarendon Press, Oxford, 1989).

\bibitem{parisi}
A.~Cavagna, I.~Giardina, and G.~Parisi, 
``Stationary points of the Thouless-Anderson-Palmer free energy",
PhysRev. {\bf B 57},11251--11257 (1998)
doi:10.1103/PhysRevB.57.11251
[arXiv:cond-mat/9710272]. 

\bibitem{example}
A.~Crisanti, L.~Leuzzi, and T.~Rizzo, ``The complexity of the spherical {p}-spin spin glass model, revisited", 
Eur. Phys. J. {\bf B 36}, 129-136 (2003) 
doi:10.1140/epjb/e2003-00325-x
[arXiv:cond-mat/0307586].

\bibitem{Witten:2010cx}
E.~Witten,
``Analytic Continuation Of Chern-Simons Theory,''
AMS/IP Stud. Adv. Math. \textbf{50}, 347-446 (2011)
[arXiv:1001.2933 [hep-th]].
   
\end{thebibliography}
\end{document}